\documentstyle[12pt]{article}
\begin{document}
\title{Quantum Statistical Processes in the Early Universe
\thanks {Invited Talk at the Conference on {\it
Quantum Physics and the Universe}
 Waseda University, Tokyo, August 1992. Proceedings edited by M. Namiki,
 K. Maeda, et al (Pergamon Press, Tokyo, 1993)}}
\author{B. L. Hu\\
{\small  Department of Physics, University of Maryland, College Park,
MD 20742, USA}}
\date{}
\maketitle
\centerline{(umdpp 93-59)}
\begin{abstract}
We show how the concept of quantum open system and the methods in
non-equilibrium statistical mechanics can be usefully applied to
studies of quantum statistical processes in the early universe.
We first sketch how noise, fluctuation, dissipation and decoherence
processes arise in a wide range of cosmological problems. We then focus
on the origin and nature of noise in quantum fields
and spacetime dynamics. We introduce the concept of geometrodynamic
noise and suggest a statistical mechanical definition of gravitational
entropy. We end with a brief discussion of the theoretical
appropriateness to view the physical universe as an open system.
\end{abstract}
\newpage

\section{Introduction and  Overview}

The aim of this talk is to show how the concept of quantum open systems
and the methods in non-equilibrium statistical mechanics can be
usefully applied to the study of quantum statistical
processes in the early universe. By the early universe
we refer to the period from the Planck time ($10^{-43} sec$) down to, say,
the GUT time ($10^{-34} sec$), a period depicted by semiclassical
cosmology. Semiclassical gravity theories assume
that the gravitational field can be treated as classical while the
matter fields are quantized. Semiclassical cosmology refers to those
classes of universes which are solutions to the semiclassical Einstein
equations
with backreaction ( the sources are usually taken to be
the vacuum expectation values of the
energy momentum tensor of the matter quantum fields) \cite{BirDav}.
It includes  many versions of inflationary
cosmology where the universe is driven by a quantum vacuum
\cite{inflation} or stochastic source \cite{StoInf}.
Prior to the Planck time is the realm of quantum cosmology \cite{qc},
where the gravitational field has also to be quantized.

The {\it key concept} we use is that of a quantum open system.
The {\it paradigm} we adopt is that of a quantum Brownian model
in a general environment.
The {\it method} we use is the Feynman-Vernon influence functional
technique.

The {\it key processes} we study are noise and fluctuation on the one hand,
and dissipation and diffusion on the other.
On the level of the influence functional one can obtain
explicit forms of the noise and dissipation kernels corresponding to different
types of environment and system-environment couplings; and from the
stochastic equations of motion (the master equation, Fokker-Planck equation,
or Langevin equation) derivable from the influence action
one can examine the processes of dissipation and diffusion at work.
Notice that on the level of noise and fluctuation, the description is
microscopic and stochastic, whereas on the level of dissipation and diffusion
the description is macroscopic and deterministic. Indeed it is the
degradation of  information in the environment which engenders the dissipative
dynamics of the system. The three steps
-- separation of system and environment,
coarse-graining the environment and the backreaction of the environment
on the system-- turns the reversible dynamics of a closed system into the
irreversible dynamics of an open system, an elemental theory into an effective
theory.  Many common issues in theoretical physics which interpolate
between the microscopic and the macroscopic, the stochastic and the
deterministic, the random and the structured, can be studied with such
a conceptual scheme and paradigm. Some examples of current interest are:

a) Relation of quantum and thermal fluctuations \cite{HuZhaUncer,CalHuFluct}.
This occurs in e.g.,
gravitational wave detectors, early universe phase transitions,
black hole radiance.

b) Quantum to classical transition \cite{decoherence,consistent,decQC}.
The conditions of
decoherence (towards consistent histories) and correlation
(peaking in phase space) leading to
a classical description are related to the diffusion process as a consequence
of noise.

c) Emergence of persistent structures in nonlinear dissipative systems.
\cite{GelHar2,BalVen,Woo}.\\

Examples of quantum processes  in semiclassical
and quantum cosmology which require non-equilibrium statistical
field-theoretica
   l
considerations are:

1. Galaxy formation from primordial quantum fluctuations

2. Entropy generation from quantum stochastic and kinetic processes

3. Phase transitions in the early universe as vacuum or noise-induced processes

4. Anisotropy dissipation from particle creation and other backreaction
processe
   s

5. Dissipation in quantum cosmology and the issue of the initial state

6. Validity of the minisuperspace approximation as a backreaction problem

7. Decoherence, backreaction  and the semiclassical limit of quantum gravity

8. `Birth' of the universe as a spacetime fluctuation and tunneling phenomenon

9. Topology change and quantum coherence problems

10. Gravitional entropy, singularity  and time asymmetry\\

As a general introduction to these ideas, a range of topics was covered
in my talk, but at varying details:

A. Introduction--Issues, Concepts and Methodology

B. Noise, Fluctuation and Structure in Inflationary Universe

C. Dissipation and Backreaction in Semiclassical Cosmology

D. Quantum Cosmology in the Light of Statistical Mechanics

E. Spacetime Noise and Gravitational Entropy

F. Discussion\\

What follows is a progress report of recent work on these problems carried
out by Calzetta, Paz,
Sinha, Zhang and myself. It is not meant to  be a review of all
the work done in this field, as there are many.
Amongst researchers in quantum cosmology there is no consensus on the
interpretation of conceptual issues.
The viewpoint I take here is probably as subjective as others'.

Since I have discussed some of these topics in general terms in other
occasions, where written reports are available, I do not want to overburden
this report here.
I would rather focus on one or two themes and discuss some new ideas,
specifically, on Topic E above.
Let me, however, mention where discussions of some of these topics can be found
in our previous work:

The adoption of open-system concepts to quantum cosmological processes
naturally emerged on the one hand from our earlier systematic studies of
particle creation backreaction problems in semiclassical cosmology
\cite{CalHu87}, and on the other from seeking a quantum field-theretical
formulation of kinetic theory \cite{CalHu88}.
A statistical mechanical meaning of dissipation in the  cosmological
backreaction problem was explored and a fluctuation- dissipation
relation for quantum field processes in a dynamical spacetime was
proposed in \cite{HuPhysica}. The appearance of dissipation in the equations
of quantum cosmology (when certain sectors of superspace are coarse-grained,
as in the minisuperspace approximation)
was brought up in \cite {CalHu89,CalDisQC,HuErice,SinHu91,HPS}
and its implications
on the specificity of the initial conditions are discussed.

Dissipation is only half of the story. In the closed-time-path (CTP) or
Schwinger- Keldysh effective action \cite{ctp,ctpcst}
we used to describe these processes
it comes from the real part only. The imaginary part which we obtained
(e. g., in the problem of a scalar field in Bianchi Type-I universe
\cite{CalHu87})
depicts the noise. We did not quite appreciate its physical meaning
until we saw it
in the context of the Feynman-Vernon influence functional description of
quantum Brownian motion. (This is a problem studied by many people, notably
\cite{FeyVer,CalLeg83,Grabert}). To see this, the framework has to be
extended from the treatment of wave functions to density matrices
(more precisely, reduced density matrices for open systems).
Formally, an intermediate step which leads
from the effective action which incorporates vacuum fluctuations
to the influence action which incorporates the averaged
effects of an environment is
the so-called coarse-grained effective action \cite{cgea}. Indeed
the in-in CGEA turns out to be just the influence action applied to vacuum
states.

It is the noise kernel which is responsible
for processes like decoherence. Studies of decoherence in quantum mechanics
and quantum cosmology have seen some interesting development from the early
80's (see talks by Zurek and Hartle in this Conference).
Structually, dissipation (relaxation)
and decoherence (diffusion) are two interelated processes both induced
by the noise and fluctuations in the environment.
For us it was from the recognition
that the closed-time-path effective action (which subtains our investigations
of the cosmological backreaction problem) is equivalent to the influence
action (which is used for the derivation of the stochastic equations) that we
confirmed our earlier speculation on the statistical mechanical meaning of
these
cosmological processes. The perturbation techniques we used in the
derivation of effective actions have become immediately useful for the
treatment of noise and fluctuation in nonlinear problems,
as we will show in the
next section. This methodology was used to derive master
equations for non-linear couplings \cite{HPZ1,HPZ2}
and generalized
fluctuation-dissipation relations for quantum Brownian motions in a
general environment in \cite{HPZ2,ZhangPhD,FDR}.

A unified viewpoint was advocated for studying the statistical aspects
of quantum cosmology in \cite{HuTsukuba},
where it is urged that the statistical processes of decoherence and
correlation,
dissipation and backreaction, noise and fluctuation be studied
in an integrated and interconnected way, not in isolation, for the discussion
of
    quantum
to classical transition, semiclassical limits, and related issues.
(Anderson \cite{And}, and Habib and Laflamme \cite{HabLaf}
first noticed the importance of decoherence
in the WKB Wigner function as an indication of correlation in phase space,
Paz and Sinha \cite{PazSin1,PazSin2} showed how a decohered WKB branch
obeys the semiclassical Einstein equations.)
The application of statistical mechanical ideas and methods for such
studies in quantum cosmology, including the minisuperspace approximation
problem mentioned above \cite{SinHu91} is described in greater details
in \cite{SinhaPhD,HPS}.
The study of how fluctuation and dissipation processes relate to
irreversibility in cosmology is discussed in \cite{HuSpain}.

As expressed in \cite{HuTsukuba}, while many of these statistical processes
have been explored in some detail in their respective contexts, the study
of noise which underlies these processes is largely untouched, at least
in the context of cosmology.  Extending the treatment of quantum Brownian
motion \cite{HPZ1,HPZ2,ZhangPhD}
to quantum fields in Minkowsky \cite{HPZ3} and then to de Sitter spacetimes
\cite{HPR},
we have recently begun an investigation of noise, fluctuation and structure in
the inflationary universe \cite{Belgium}.
We showed from first principles how one can
derive classical stochastic dynamics from quantum field theory
(in curved spacetime, if for cosmological purposes).
We showed how the coarse-graining of  an environment field
(e.g., in Starobinsky's stochastic inflation model \cite{StoInf},
it is  the high
frequency modes inside the de Sitter horizon) leads to noise and
dissipation in the (functional) stochastic equations.
With these equations one can study how readily the
different sectors of the spectrum decohere, and check on the validity
of the commonly invoked assumption that the long wavelength modes behave
classically. In this setting, we also proposed a new mechanism of
colored noise generation from nonlinear coupling of the inflation field
with the bath. These results are detailed in \cite{Belgium}.
Colored noise  would give rise to  non-Gaussian fluctuations.
The implication of these effects on galaxy formation is under investigation
\cite{HP}.
The above is a sketch of topics  A to D of my talk.
In the following I will focus on the ideas in topic E, i.e.,
origin of noise in quantum fields and spacetime dynamics, and suggest
an alternative definition of gravitational entropy. We shall examine
the soundness of viewing the physical universe as a quantum open system
in the Discussion.

\section{ Noise from Coarse-Grained Interacting Quantum Fields}

We have worked out four examples, one in field theory: 1) a $\lambda \phi^4$
theory where two self-interacting scalar fields are coupled bi-quadratically
\cite{HPZ3};
one for semiclassical cosmology: 2) a free scalar field propagating in a
Bianchi
Type I background spacetime \cite{CalHu87}; and two in quantum cosmology:
3) a minisuperspace consisting of the Robertson-Walker (RW)
scale factor and the lowest mode of a $\lambda \phi^4$ scalar field is coupled
to the higher scalar modes \cite{SinHu91};
and 4) the fully quantum version of the scalar field in Bianchi Type I (BI)
universe problem \cite{PazSin2}. The two quantum cosmological examples
have the same coupling types as the two corresponding problems in field theory
and semiclassical cosmology, therefore we shall use these two examples as
models to show how noise can be associated with the coarse-graining of
quantum fields and spacetimes.

In \cite{HPZ2} we used a quantum Brownian model where a system (of one
harmonic oscillator) is coupled nonlinearly to a bath (of many oscillators)
to show how one can obtain different types of colored noise from different
couplings (we analyzed a polynomial type) and from different baths
(we analyzed both ohmic and nonohmic spectral densities).
In extending this work to field theory \cite{HPZ3}, we discussed in particular
t
   he
case of bi-quadratic coupling, and a zero temperature bath.
Hence the noise is originated from vacuum fluctuations.
This brings the present statistical field problem even closer to that of
the quantum field problem studied before via CTP effective actions.

Consider two independent self-interacting scalar fields:
$ \phi(x) $ depicting the system, and $ \psi(x) $ depicting the bath.
The classical action for these two fields are given respectively by:

\begin{equation}
S[\phi]
=\int d^4x~\Bigl\{
  {1\over 2}\partial_{\nu}\phi(x)\partial^{\nu}\phi(x)
 -{1\over 2}m^2_{\phi}\phi^2(x)
 -{1\over 4!}\lambda_{\phi}\phi^4(x) \Bigr\}
\end{equation}

\begin{equation}
S[\psi]
=\int d^4x~\Bigl\{
  {1\over 2}\partial_{\mu}\psi(x)\partial^{\mu}\psi(x)
 -{1\over 2}m^2_{\psi}\psi^2(x)
 -{1\over 4!}\lambda_{\psi}\psi^4(x) \Bigr\}
=S_0[\psi]+S_I[\psi]
\end{equation}

\noindent where $ m_{\phi} $ and $ m_{\psi} $ are the bare masses of
$ \phi(x) $ and $ \psi(x) $ fields respectively. Both fields have a
quartic self-interaction with the bare
coupling constants $ \lambda_{\phi} $ and $ \lambda_{\psi} $.
In (2.2) we have written $S[\psi]$ in terms of a free part $S_0$ and
an interacting part $S_I$ which contains $ \lambda_{\psi}$.
Assume these two scalar fields interact via a bi-quadratic coupling

\begin{equation}
S_{int}
=\int d^4x~\Bigl\{
 -\lambda_{\phi\psi}\phi^2(x)\psi^2(x) \Bigr\}
\end{equation}

\noindent Assume also that all three coupling constants
$ \lambda_{\phi} $, $ \lambda_{\psi} $ and $ \lambda_{\phi\psi} $
are small parameters of the same order.
The total classical action of the combined system plus bath field is
then given by

\begin{equation}
S[\phi,\psi]
=S[\phi]+S[\psi]+S_{int}[\phi,\psi]
\end{equation}

\noindent The total density matrix of the combined system plus bath field is
defined by
\begin{equation}
\rho[\phi_f,\psi_f,\phi'_i,\psi'_f,t]
=<\phi_f,\psi_f|~\hat\rho(t)~|\phi'_i,\psi'_f>
\end{equation}

\noindent where $ |\phi> $ and $ |\psi> $ are the eigenstates of the field
operators $ \hat\phi(x) $ and $ \hat\psi(x) $,

\begin{equation}
\hat\phi(\vec x) |\phi>=\phi(\vec x) |\phi>, ~~~
\hat\psi(\vec x) |\psi> = \psi(\vec x) |\psi>
\end{equation}

Since we are interested in the behavior of the system, and the environment
only to the extent of how it influences the system, the quantity of
relevance is the reduced density matrix defined by

\begin{equation}
\rho_r[ \phi, \phi', t] \equiv \int d \psi \int d \psi' \delta (\psi-\psi')
\rho [\phi, \psi, \phi', \psi',t]
\end{equation}

\noindent Here $~\int d\psi(\vec s)~$ denotes the functional
integral over the Hilbert space of all quantum states of the $ \psi $ field.
For technical convenience, let us assume that the total density matrix
at an initial time is factorized, i.e.,
that the system and bath are statistically independent,
\begin{equation}
\hat\rho(t_0)
=\hat\rho_{\phi}(t_0)\times\hat\rho_{\psi}(t_0)
\end{equation}

\noindent where $ \hat\rho_\phi(t_0) $ and $ \hat\rho_{\psi}(t_0) $ are the
init
   ial
density matrix operator of the $ \phi $ and $ \psi $ field respectively,
the former being equal to the reduced density matrix $\hat \rho_r$ at $t_0$
by this assumption.

The influence functional $ ~F[\phi,\phi']~ $ which summarizes the averaged
effect of the bath on the system is defined as
$$
F[\phi,\phi']=
  \int d\psi_f(\vec x)
   \int d\psi_i(\vec x)
   \int d\psi'_i(\vec x)~
   \rho_{\psi}[\psi_i,\psi'_i,t_0]~
   \int\limits_{\psi_i(\vec x)}^{\psi_f(\vec x)}D\psi
   \int\limits_{\psi'_i(\vec x)}^{\psi_f(\vec x)}D\psi'
$$
\begin{equation}
  \times\exp i\Bigl\{S[\psi]+S_{int}[\phi,\psi]
    -S[\psi']-S_{int}[\phi',\psi'] \Bigr\}
\end{equation}

\noindent Here $~\int D\psi~$
denotes the functional path integral over all possible
histories of the $ \psi $ field under some boundary conditions.
The influence action $ \delta A[\phi,\phi'] $
and the effective action $ A[\phi,\phi'] $ are defined as

\begin{equation}
F[\phi,\phi']=\exp i\delta A[\phi,\phi']; ~~~
A[\phi,\phi']
=S[\phi]-S[\phi']+\delta A[\phi,\phi']
\end{equation}

The quantum average of a physical variable $Q[\psi, \psi']$
over the unperturbed action $ S_0[\psi] $ can be expressed in terms of
$ F^{(1)}[J_1,J_2] $, the
influence functional of the free bath field, assumed to be linearly coupled
with external sources $J_1$ and $J_2$.
For a zero temperature bath, the bath field $ \psi $ is in a vacuum state,
\begin{equation}
\hat\rho_\psi(t_0)=|0><0|
\end{equation}
\noindent then the influence functional $F^{(1)}[J_1,J_2]$ is the same as
the closed-time-path (CTP) or `in-in' vacuum generating functional
\cite{ctp,ctpcst} and the associated influence action
is the usual CTP or in-in vacuum effective action. In such cases, the
two-point functions of the bath fields
are just the well-known Feynman, Dyson and positive-frequency
Wightman propagators for a free scalar field.

If $ \lambda_{\phi\psi} $ and $ \lambda_{\psi} $ are assumed to be small
parameters, the influence functional can be calculated perturbatively
by making a power expansion of $ \exp i\bigl\{S_{int}+S_I\bigr\} $
in orders of $\lambda$ and $\hbar$. This is similar to
the perturbation calculation for $\lambda \phi^4$ theory in the CTP formalism
carried out before for quantum fluctuations \cite{CalHu87}
and for coarsed-grained fields \cite{cgea}.
The 1-loop effective action up to $O(\lambda^2)$ is \cite{ZhangPhD,HPZ3}:

\begin{eqnarray*}
A[\phi,\phi']
& = & \Bigl\{S[\phi]+\delta S_1[\phi]+\delta_2[\phi]\Bigr\}
  -\Bigl\{S[\phi']+\delta S_1[\phi']+\delta_2[\phi']\Bigr\}
  +\delta A[\phi,\phi'] \\
& = & S_{ren}[\phi]+\int d^4x\int d^4y~{1\over 2}
   \lambda^2_{\phi\psi}\phi^2(x)V(x-y)\phi^2(y) \\
&&  -S_{ren}[\phi']-\int d^4x\int d^4y~{1\over 2}
   \lambda^2_{\phi\psi}\phi'^2(x)V(x-y)\phi'^2(y)\\
\end{eqnarray*}
$$
 -\int\limits_{t_0}^tds_x\int d^3\vec x
   \int\limits_{t_0}^{s_y}ds_y\int d^3\vec y~
   \lambda^2_{\phi\psi}
   \Bigl[\phi^2(x)-\phi'^2(x)\Bigr]
   \times\eta(x-x')  \Bigl[\phi^2(y)+\phi'^2(y)\Bigr]
$$
\begin{equation}
+i\int\limits_{t_0}^tds_x\int d^3\vec x
   \int\limits_{t_0}^{s_x}ds_y\int d^3\vec y~
   \lambda^2_{\phi\psi}
   \Bigl[\phi^2(x)-\phi'^2(x)\Bigr]
   \times \nu(x-y)   \Bigl[\phi^2(y)-\phi'^2(y)\Bigr]
\end{equation}
\noindent where $ S_{ren}[\phi] $ is the renormalized action of the $\phi $
fiel
   d,
now replaced by the physical mass $ m^2_{\phi r} $ and physical coupling
constan
   t
$ \lambda_{\phi r} $, namely,
\begin{equation}
S_{ren}[\phi]
=\int d^4x\Bigl\{{1\over 2}\partial_{\mu}\phi\partial^{\mu}\phi
-{1\over 2}m^2_{\phi r}\phi^2-{1\over 4!}\lambda_{\phi r}\phi^4\Bigr\}
\end{equation}

\noindent and the kernel for the non-local potential in (2.12) is
\begin{equation}
V(x-y)
=\mu(x-y)-sgn(s_x-s_y)\eta(x-y)
\end{equation}
\noindent which is symmetric and $\eta$, $\nu$, $\mu$ are real and nonlocal
kernels:

\begin{equation}
\eta(x-y)
={1\over 16\pi^2}\int {d^4p\over (2\pi)^4}~e^{ip(x-y)}
{}~\pi\sqrt{1-{4m^2_{\psi}\over p^2}}~\theta(p^2-4m^2_{\psi})
 \times isgn(p_0)
\end{equation}
\begin{equation}
\nu(x-y)
={2\over 16\pi^2}\int{d^4p\over (2\pi)^4}~e^{ip(x-y)}
{}~\pi~\sqrt{1-{4m^2_{\psi}\over p^2}}~\theta(p^2-4m^2_{\psi})
\end{equation}
\begin{equation}
\mu(x-y)
=-{2\over 16\pi^2}\int{d^4p\over (2\pi)^4}~e^{ip(x-y)}
 \int\limits_0^1d\alpha\ln\Bigl|1-i\epsilon-
 \alpha(1-\alpha){p^2\over m^2_{\psi}}\Bigr|
\end{equation}

$\eta$ and $\nu$ are called the dissipation and noise kernels respectively.
This is because the real part of the influence functional is responsible for
dissipation, while its imaginary part
can be viewed as arising from a noise source $ \xi(x)$ whose
distribution functional is given by
\begin{equation}
P[\xi]
=N\times\exp\biggl\{-{1\over 2}\int d^4x \int d^4y
 \xi^2(x)~\lambda^{-2}_{\phi\psi}\nu^{-1}(x-y)~\xi^2(y)\biggr\}
\end{equation}

\noindent  where $ N $ is a normalization constant. The action describing
the noise $ \xi(x) $ and system field $ \phi(x) $ coupling is
\begin{equation}
\int d^4x~\Bigl\{~\xi(x) \phi^2(x)~\Bigr\}
\end{equation}

\noindent
In the associated functional Langevin equation for the field, the
corresponding stochastic force arising from a biquadratic coupling is
\begin{equation}
F_{\xi}(x)\sim\xi(x)\phi(x)
\end{equation}

\noindent which constitues a multiplicative noise.
 (See, e.g. \cite{stochastic}).

{}From the influence action (2.12), it is seen that the dissipation generated
in the system by this noise is of the nonlinear non-local type.
If we define the dissipation kernel $ \gamma(x-y) $ by
\begin{equation}
\eta(x-y)
={\partial\over\partial (s_x-s_y)}\gamma(x-y)
\end{equation}

\noindent then
\begin{equation}
\gamma(x-y)
={1\over 16\pi^2}\int {d^4p\over (2\pi)^4}~e^{ip(x-y)}
 \pi\sqrt{1-{4m^2_{\psi}\over p^2}}~
 \theta(p^2-4m^2_{\psi})~{1\over |p_0|}
\end{equation}

\noindent In the Langevin field equation, the dissipative force is

\begin{equation}
F_{\gamma}(x)\sim
\Biggl\{\int d^4y~\eta(x-y)\phi^2(y)~\Biggr\}\phi(x)
\end{equation}

As discussed in  \cite{ZhangPhD,HPZ3},
we find that a fluctuation-dissipation relation exists between
the dissipation kernel (2.15) and the noise kernel (2.16) :

\begin{equation}
\nu(x)=\int d^4y~K(x-y)\eta(y)
\end{equation}
\noindent where
\begin{eqnarray*}
 K(x-y)& = &\delta^3(\vec x-\vec y)  \int\limits_{-\infty}^{+\infty}
   {d\omega\over 2\pi}~ e^{i\omega(s_x-s'_x)}|\omega| \\
& = & \delta^3(\vec x-\vec y)  \int\limits_0^{+\infty}
   {d\omega\over\pi} \omega\cos\omega(s_x-s_y) \\
\end{eqnarray*}
\noindent Apart from the delta function $ \delta^3(\vec x-\vec x') $, the
convolution kernel for quantum fields has exactly the same form as for
the quantum Brownian harmonic oscillator with linear  or nonlinear
dissipations at zero temperature.

\section{ Geometrodynamic Noise}
\setcounter{equation}{0}

We now use the results of the above example to discuss the problem of noise
in quantum cosmology. In the minisuperspace model discussed by Sinha and
Hu \cite{SinHu91} (SH) the system is made up of the scale factor $a$
of a Robertson-Walker Universe and the homogeneous mode $\chi$ of a massive
($m$) interacting ($\lambda$)
scalar field. The higher inhomogeneous modes $f_k$ make up the environment.
Since the equation of
motion for the linearlized gravitational perturbations (the Lifshitz
equations) has the same form as a massless minimally coupled scalar field,
they can be used to mimic the gravitational modes customarily ignored
when one takes the minisuperspace approximation. The interaction
action is

\begin{equation}
S_{int}= -{1 \over 2}\int d\eta [m^2 a^2 f_k^2 - {\lambda \over {4 !}}
 (6 \sum _k \chi_0^2 f_k^2 + ...)]
\end{equation}

\noindent where $\eta = \int dt/a$ is the conformal time, and the dots
denote higher order $f$ terms, whose effect is ignored here.
The coupling between the system ($a, \chi_0$) and the bath ($f_k$)
is of a biquadratic kind.
The framework which Sinha and Hu used is in terms of the in-in effective action
for wave functions of the universe. To discuss noise one needs to extend
the framework to that of influence functionals dealing with
reduced density matrices constructed from the wavefunctions of the
minisuperspace. However, using prior experience in interpolating
between these two formalisms, we can almost guess the
result. For a zero temperature bath consisting of vacuum fluctuations of
the higher scalar field modes, we expect the noise kernel be the same as (2.16)
in the example of Sec. 2.
This noise we can call a minisuperspace noise, it is actually
the noise experienced in the minisuperspace sector from coarse-graining all
the other modes.

Since the SH model is meant to mimic the gravitational interactions,
let us see how this can be applied to gravitation coarse-grainings.
For a purely gravitational problem where the system consists of the
homogeneous universe (e.g., Robertson-Walker universe or Bianchi universes)
and the environment consists of the inhomogeneous modes (or perturbations of
the homogeneous background spacetime),
such as that studied by Halliwell and Hawking \cite{HalHaw},
one can deduce a noise in the minisuperspace arising from
coarse-graining the inhomogeneous gravitational
modes and call it, apppropriately, a `geometrodynamic' or `spacetime' noise.
Of course, in the case of gravitational interaction, the coupling between
the minisuperspace and the other gravitational modes is different from
the SH model. In the perturbative  case, they are of the derivative type.
\footnote{They come from separating the scalar curvature term $R$
in the Einstein action into two groups $g^0 + h^1$, the background spacetime
and the gravitational perturbations, and $R$ has two derivatives
in $g$.}

Another well-studied example which we  can use to deduce noise associated
with quantum field- and spacetime- coarse-grainings is  that of a
scalar field in a Bianchi Type I universe.
The system is the BI universe with line element
\begin{equation}
ds^2 = a^2 ( d \eta^2 -e^{2 \beta}_{ij} dx^i dx^j)
    \equiv g_{\mu \nu} dx^\mu dx^\nu
\end{equation}
\noindent where $\eta$ is the conformal time and $\beta_{ij}$ is the
(traceless)
anisotropy matrix.
The bath is made up of a massless conformal
scalar field with classical action
\begin{equation}
S_m = {1 \over 2}\int d^4x (g^{\mu \nu} \partial _\mu \phi \partial _ \nu \phi
      -{1 \over 6} R \phi^2).
\end{equation}
\noindent where $R$ is the scalar curvature.
Calzetta and Hu \cite{CalHu87} have derived the form of the Schwinger-Keldysh
effective action. The real and imaginary parts of it one can identify as
the dissipation and noise kernels.  Before, we only concentrated on the
former part as we were interested in the dissipative effects on the dynamics
of spacetime due to the backreaciton of particle creation
in the scalar field. Here we would like to explore the noise aspect.
The extension to the framework of reduced density matrices
has been carried out by Paz and Sinha \cite{PazSin2}.
They used this model in quantum cosmology
to illustrate the effects of decoherence, correlation
and backreaction for attaining the semiclassical limit.
One can use the QBM paradigm illustrated in the previous section to
treat this problem in quantum cosmology. Paz and Sinha derived
the influence functional for the BI universe \cite{PazSin2}.
{}From it one can examine the noise kernel and derive the distribution
function associated with the noise force. Details of this investigation
can be found in \cite{HuSin}.

The noise in this model of course originates from the scalar field.
But, by the same reasoning, one can construct similar models where
the environment is of gravitational origin. Gravitational perturbations of a
BI universe is an example. (The classical backreaction problem was studied by
Hu \cite{Hu78}). The Gowdy universe \cite{Gowdy}
is another interesting model where
the system can be taken to be the homogeneous mode and the environment
the inhomogeneous modes. One can calculate the form of noise associated with
the gravitational wave modes.
Sinha and I are in the process of calculating the geometrodynamic noises
for these quantum cosmological models ( e.g., the minisuperspace RW models
\cite{SinHu91}, the Bianchi I model \cite{PazSin2},
the midi-superspace model of perturbed RW universe \cite{HalHaw},
and the Gowdy models).

\section{ Gravitational Entropy}
\setcounter{equation}{0}

In the general framework of statistical field theory as exemplified by the
quantum Brownian model discussed above it is natural to
define an entropy associated with the reduced density matrix of the
system with the coarse-grained effect of the environment taken into
consideration. It is given formally by
\begin{equation}
S \equiv - Tr~\rho_{red} ln \rho_{red}
\end{equation}

For the cases where the environment is a coarse-grained
quantum field, this can be regarded as the
entropy of a quantum field; whereas for the cases
where the environment is a coarse-grained spacetime in the form
of inhomogeneous cosmological modes or gravitational waves,
this can be rightfully called spacetime or gravitational entropy.
(For a related definition see \cite{HuKan,KanCQG}.)
It is a well-known
result in statistical mechanics that any
entropy function associated with a closed system remains
constant in time but the entropy functions constructed from the reduced
density matrix of a subsystem with coarse-graining in its environment
increases with time.
For quantum cosmology, there is no intrinsic time. Operationlly, time is
a parameter chosen to impart `dynamics' for the evolution of the system.
One can use the WKB time associated with the semiclassical
limit. The direction of time associated with the expansion of the
universe defines a cosmological arrow of time.
(Often in quantum cosmology the volume of the universe is chosen as the time
parameter.)
Information in an open system is degraded by the environment and
flows in the direction of entropy increase. The increase of entropy
(with respect to this cosmological time) defines a thermodynamic  arrow of
time.
\footnote{Since we are not dealing with ordinary matter, but with
coarse-grainings in spacetimes, perhaps one should be careful in  using the
word
thermodynamic here: it need not refer to matter, even though the high-frequency
gravitational waves can indeed be viewed as a form of matter.}

Let me first distinguish the meaning of gravitational entropy defined
in our present context from that used previously in different contexts.
Penrose \cite{Penrose}
first introduced this concept as a way to describe how many distinguishable
configurations or degrees  of freedom a gravitational field or
spacetime geometry has. He suggested that some
intergrated form of the square of the Weyl curvature tensor $C_{\alpha
\beta \gamma \delta}C^{\alpha \beta \gamma \delta}$ can serve as
a  measure of how `coarse' or `irregular' the spacetime is. This is a
reasonable suggestion as the Weyl curvature measures the gravitational
wave components of spacetime
(e.g., the RW universe which is isotropic and homogeneous has $C=0$.)
Penrose believes that the universe begins in a highly regular gravitational
state, and ends with a highly irregular one.
He proposed this Weyl Curvature Hypothesis, i.e., the increase of gravitational
entropy proportional to the square of the Weyl curvature,
to describe the increase of observed gravitational clumping
as the universe expands.

Hu \cite{Hu83} discussed gravitational entropy in the extended context
when quantum field processes like particle creation are important.
He brought in the effect of backreaction of
quantum matter on the dynamics of geometry and
showed that if one adopts Penrose's geometric notion of gravitational
entropy, particle creation in an anisotropic or inhomogeneous universe
(which is proportional to the square of the Weyl curvature tensor) can act
as an efficient  process near the Planck time to `convert'
gravitational entropy (associated with spacetime) to
matter entropy (associated with the created particles).

The gravitational entropy defined here in a statistical mechanical sense is
very different from Penrose's geometric definition both in terms of
syntex and context. Choosing the Weyl curvature as  a measure of
the entropy of a gravitational field is a descriptive definition but it lacks a
sound theoretical basis.
\footnote{The underlying logic runs somewhat like:
1) One knows from second law of
thermodynamics that entropy increases with time  ( does it apply equally well
to entropy of gravitational field?); 2) History of the universe
shows (rather, one believes) that spacetime evolves from
a highly regular state to a highly irregular state, the
Weyl tensor increasing in time correspondingly; 3) Hence pick the Weyl tensor
(squared) as a definition of gravitational entropy.}
By contrast the present definition
of gravitational entropy adheres to the formal premises of (non-equilibrium)
statistical mechanics. But one wants to know how relevant it is in
describing reality?--How good is it in helping us understand the nature of
physical laws involving gravity,  and the behavior of our  universe ?
The questions we need to address are thus two-fold:
1) the physical meaning of such a definition and its relation
with the more intuitive and descriptive definition of Penrose.
2) Since the definition of entropy here depends on how the reduced
density matrix is constructed, and hence on how the system is
defined with respect to its environment, a deeper question to ask is:
`Who'   decides what should be called the system and how is this decision
justified?
The first issue relates to the specific context of gravity and cosmology,
the second issue points to the basic tenets of statistical mechanics.
We will have space here only to mention some ideas on these important
problems. Detailed discussions are to be given elsewhere.

Despite the difference in syntex from which these definitions are introduced,
and the difference in emphasis they attach--
Penrose's (79) definition is of a geometrical
nature, Hu's (83) incorporates quantum fields, and this one (93) is based on
statistical mechanics concepts-- they are nevertheless interelated with each
other. Hu's (83) is related to Penrose's in that the Weyl tensor which measures
the `disorder' of geometry in Penrose's definition of gravitational entropy
also measures the amount of particle creation in the field; and it is due to
the backreaction of created particles that gravitational entropy changes
(indeed, decreases).
Now, the present definition uses the same but extended
theoretical framework as the Hu's 1983 work, as we explained above, so they
are consistent with each other.
The essential physics described in both places lies in the dissipation and
noise kernels. The 83 paper focuses on the former aspect while here is added
the latter aspect.  For quantum processes in anisotropic and inhomogeneous
spacetimes, both are proportional to the Weyl curvature tensor (squared).
Thus, in the conception of \cite{HuPhysica}, the noise associated with
vacuum fluctuations of the field (or spacetime) when parametrically amplified
gives rise to particle creation which is measured by the Weyl curvature
squared, while the dissipative term in the Langevin equation for the
probability distribution of the geometry variables,
which measures in this capacity the damping of anisotropy in the dynamics of
spacetime, is also proportional to $C^2$. This is not surprising
as the noise and dissipation are connected by a fluctuation -dissipation
relation \cite{HPZ2,HuSin}. (See \cite{CanSci,Mot} for such relations in
spacetimes with event horizons.)
 Indeed, as emphasized in \cite{HuPhysica},
the backreaction problem in cosmology, i.e., the mutual influence
between quantum matter in the form of particle creation and
geometry through the laws of geometrodynamics,  can be seen
as a manifestation of a generalized fluctuation-dissipation relation
between matter and spacetime.

For other discussions of gravitational entropy, see \cite{graventropy}.


\section{The Physical Universe as a Quantum Open System}

To close, let me now say a few words about the second issue above,
i.e., how appropriate
is it to view the universe as an open system and to what extent can the
basic tenets of non-equilibrium statistical mechanics be applied to
cosmological problems. In describing some physical entity as an open system
the criteria in the separation of the system and environment and the choice
of coarse-graining enter in a fundamental way.
One can ask the following questions:

1) How natural?

In cosmology many physical problems of interest
have `natural' separations of system and environment
-- event horizon, particle horizon, causal boundaries,
multiply connectedness, etc. So an open system description of
cosmological problems is not just an artificial scheme, but
{\it can} be meaningful. (See last section of \cite{HuKan}.)
However, how `natural' the separation is can vary with the degree of
accuracy (fineness of description) or level of structure  (hierarchy of
compositeness) one restricts one's attention to.
(See last section of \cite{HuSpain}.)
For example, a background field or spacetime separation gives meaningful
description in the semiclassical regime, but can become meaningless
as the interaction between matter and gravitational degrees of freedom
becomes highly nonlinear or nonlocal.
A black hole in the full quantum gravity regime
may evaporate completely,
and with the disappearance of the event horizon, the semiclassical
description premised upon such a separation is no longer meaningful.
But at each definite level it is permissible to consider the system as
open. In this sense a theoretically closed system which requires a large
amount of detailed information for its complete description can in
principle be approximated successively by a hierarchy of ordered open systems.

2) How open?

If, say, an event horizon used for the separation of the system and
environment changes with time for some observer,
does it mean that the system sees different things at different moments,
as the system and noise both change. The answer is yes, but it is really
not so alarming.
(This is indeed the case underlying the workings of `stochastic inflation'
in the de Sitter universe,
where the higher modes, which make up the noise, leave
the horizon at every moment.)
   .
As for the particle horizon,
this is happening to our physically observable (causally-connected)
universe at every moment. Our ever widening horizon is no cause for concern
even though the treatment of quantum fields with moving boundaries
(changing constituents in the Hilbert space ) is not exactly easy.

3) How sensitive is the behavior of the system to changes in the environment
and variations of coarse-grainings?

Is coarse-graining a subjective or objective effect?
\cite{BalVen}
Are there particular sets of coarse-graining which can lead to stable
or persistent structures? \cite{GelHar2}
It is believed that that part of
the classical world which possesses a definite structure
should be largely independent of the choice and influence of its surroundings.
This is the case when the system is, say, in its hydrodynamic
(long wavelength, low energy) regime. In the linear coupling regime
microscopic derivations of the transport coefficients are also largely
independent of variations in the environment \cite{FeyVer}.
These are familiar examples of
the conditions upon which robust structures can arise.
This basic issue in quantum and statistical mechanics
certainly merits more serious studies \cite{HuSpain}.

4) How good is the choice of the homogeneous cosmology as the system and
viewing the inhomogeneous sector and the matter fields as the environment?

This is, of course, more often than not, a matter of convenience rather
than principle. However, one may ponder why the physical universe is indeed
adequately described by a homogeneous cosmology in a theoretical sense.
This has been attempted in a few directions before, such as: \\
a) viewing the prominance of the homogeneous modes as a result of
infrared dominance and dimensional reduction, in a sense similar to the
Kaluza-Klein idea. This was proposed in \cite{HuErice}. \\
b) viewing the gravitational excitations as collective modes.
The distinction between the prominant modes which consititute the `system'
and the others (inhomogeneous modes) which are relegated to the `environment'
may be determined at a higher level in the hierarchy of spacetime structures.
(For example, the rotational
collective modes of the nucleus  are not fundamental variables, but they are
more appropriate in the description of macroscopic motions of the nucleus
than quarks and gluons.
An open system model of the collective variables can adequately describe
the dissipative dynamics for the nucleus -- it is impractical
and unnecessary to invoke the details of the nucleon wavefunctions,
but this will be a wrong model to use if one wants to probe into the
nuclear effects of
quark-gluon interactions.) The choice of an open system is not
arbitrary, but should be guided by the physical conditions which defines
the problem one wishes to study. More discussions of these issues can be
found in the last section of \cite{HuSpain}.\\

{\bf Acknowledgements} ~~It is a pleasure to thank Prof. Maeda for making the
arrangements for me to visit Waseda University and attend this interesting
Conference. It is probably the first one of its kind where researchers from
three basic disciplines of physics join force and compare notes together.
My visit to Japan is supported by a Senior Fellowship of the
Japan Society for the Promotion of Science.
The hospitality of Professors K. Maeda, H. Sato and K. Sato is warmly
appreciated.
The work reported in Sec 3 is carried out in collaboration with
Sukanya Sinha, with whom I have shared many enjoyable discussions.
Research is supported in part by the National Science Foundation under
grant No. PHY91-19726.

\end{document}